\documentstyle[11pt,aas2pp4]{article}

\begin{document}
\title{A MODEL FOR THE DENSITY DISTRIBUTION OF VIRIALIZED CDM HALOS}
\author{{\sc Andreas Kull} \\Canadian Institute for Theoretical Astrophysics,\\
60 St. George Street, Toronto, M5S 1A7, Canada}

\begin{abstract}
An analytic collapse model for the formation and density distribution of virialized 
cold dark matter halos is proposed. Hierarchical structure formation is taken
into account explicitly. Monte Carlo methods are used to generate samples
of mass histories of virialized halos. The mean density distribution 
found from the collapse model is in good agreement with numerical results
in the mass range from $10^{11}M_\odot$ to $10^{15}M_\odot$ and in the 
radial range form $0.05 r_{200}$ to $r_{200}$.
\end{abstract}

\keywords{cosmology: theory --- dark matter --- galaxies: halos --- methods: analytical}

\section{Introduction}
The current standard paradigm of cosmological structure formation is summarized
by three main points: The Universe on large scales is dominated by collisionless 
dark matter; the seeds of collapsed systems are small-amplitude Gaussian fluctuations 
of the uniform primordial density field and the evolution towards virialized 
systems occurs by gravitational instability. In hierarchical models of structure 
formation, the amplitude of the primordial density fluctuations increases with 
decreasing scale. This leads to a picture where small objects form first and, by 
subsequent merging, form larger objects later on. Numerical simulations 
(e.g. Lacey \& Cole 1994) confirmed this picture to a large extent. 

A striking outcome of numerical simulations is the `universal' shape of the
profile of virialized density distributions e.g. of the form
(Navarro et al. 1996, NFW96 hereafter)
\begin{equation}  
\rho(r)={\rho_0\over{(r/r_0) [1+(r/r_0)]^{2}}}
\label{nfwprof}
\end{equation}
where $r_0$ is a characteristic length scale and $\rho_0$ a characteristic 
density. Recent numerical simulations (e.g. Moore et al. 1998) indicate 
a somewhat steeper innermost slope $\rho \sim r^{-1.4}$. However, in the
radial range $[0.05 r_{200}, r_{200}]$ a broad agreement is found.  

Early analytical studies of the density distribution of collapsed 
objects are based on spherically symmetric systems 
(Gunn \& Gott 1972, Gunn 1977) forming around sharp density peaks. 
Gaussian initial conditions have been taken into account by 
Hoffman \& Shaham (1985). These studies suggest a virialized density
distribution of the form $\rho \propto r^{-\gamma}$ with
$\gamma \sim 2$. From the point of view of the current standard paradigm of structure 
formation these models suffer from one main defect: They assume
continuous and self-similar infall. There have been several attempts 
to develop more consistent models. Most of these studies seek to explain the
form of the density distribution (\ref{nfwprof}). For example, it has been 
argued (Syer \& White 1998) that it arises from repeated mergers. 
Semi-analytical (Avila-Reese et al. 1998) and combined
analytical and numerical approaches have been conducted as 
well. By including dynamical friction and tidal stripping, 
Nusser \& Sheth (1998) e.g. find that the final density profile
mainly depends on the rate of infall. The scaling of the density
distribution (\ref{nfwprof}) with formation redshift has been
investigated by Raig et al. (1998).

One of the key issues regarding hierarchical structure formation
is how well the formation history is remembered by a collapsed
system. The aim of this letter is to investigate this question.
As a working hypothesis we assume that each step of the formation 
history is remembered by the conservation of the corresponding 
total energy of infalling matter. Under this assumption, a simple 
spherical symmetric collapse model is proposed. The model is 
developed along the line of earlier analytical studies 
(e.g. Gunn 1977, Hoffman \& Shaham 1985). 
However, the condition of self-similarity is abandoned and the 
hierarchical formation history is take into account explicitly.
%The mean density distribution of dark matter halos in the 
%mass range from $10^{11}M_\odot$ to $10^{15}M_\odot$ and in the 
%radial range form $0.05 r_{200}$ to $r_{200}$ is in good
%agreement with the numerical results of NFW96. 

\section{The Model}\label{model}
The model is based on a) formation history
and  b) a collapse model describing how infalling matter virializes. The 
cosmological context is implicitly present in a). 

\subsection{Formation History\label{history}}
The excursion set formalism (Bond et al. 1991, Lacey \& Cole 1993, 1994) 
provides an analytical model of the merger history of a collapsed object. 
Given the mass $M_k$ of an object at redshift $z_k$, it predicts the 
probability distribution of the mass $M_{k+1}$ of the progenitor at
an earlier epoch $z_{k+1}$. Monte Carlo methods can be used to
generate particular merger histories. In doing so, we follow here
the method of Eisenstein \& Loeb (1996).
Starting from a halo mass $M_0$ at $z_0=0$, the merger history is traced
by stepping backwards in time until a typical mass marking the seeding 
of the object is reached. A particular mass history then is described 
by a set $\{(m_k,z_k),$ $k=0,1,2,\ldots,K\}$ where $m_k=M_{k}-M_{k+1}$ 
is the mass falling in at redshift $z_k$ and $K$ denotes the total 
number of formation steps.

\subsection{Total Energy of Infalling Matter}
In the spherical symmetric case, the total
energy $E_k$, the infalling mass $m_k$ and its turnaround radius 
$R_m$ are related by
\begin{equation}
E_k=-{G m_k M_{k+1} \over{R_m}}
\label{etot1}
\end{equation}
where $M_{k+1}$ is the mass interior of $R_m$, i.e. the mass collapsed at $z>z_k$.
For $m_k \ll M_{k+1}$, $R_m$ is approximately
\begin{equation}
R_m=\left({3 M_{k+1} \over{4 \pi \rho_m}}\right)^{1/3}
\label{rm}
\end{equation}
where $\rho_m$ is the mean density of the region of mass $M_{k+1}$ at maximum expansion.
For an Einstein-de Sitter cosmology ($\Omega=1$, $\Lambda=0$) considered here 
(e.g. Peebles 1980)
\begin{equation}
\rho_m={9 \pi^2 \bar{\rho}_0 (1+z_{k+1})^3 \over{16}}.
\label{enh2}
\end{equation} 
Here $\bar{\rho}_0$ is the background density at the current epoch $z=0$
and $z_{k+1}$ is the redshift at collapse.
From eqs. (\ref{etot1}) and (\ref{enh2}) the total energy $E_k$ of matter 
of mass $m_k$ falling in at formation step $k$ is
\begin{equation}
E_k=-(3/4)^{1/3} \pi G m_k M_{k+1}^{2/3} \bar{\rho_0}^{1/3} (1+z_{k+1}).
\label{etot}
\end{equation}
How does this approximation of $E_k$ fit the general 
picture of hierarchical structure formation? The basic assumption which has been 
made is that the density distribution of collapsed matter is approximately 
spherical symmetric, an approximation which is reasonable for virialized systems. 
No assumption about the distribution of the infalling matter has been 
made, however. The infalling matter can be distributed continuously or
it can be concentrated in one or several spot at $r=R_m$.
In this sense the approximation for $E_k$ accounts for a hierarchical structure 
formation process where larger objects form by the collapse of already 
collapsed smaller objects.

\subsection{Collapse Model\label{cmodel}}
The working hypothesis for the proposed collapse model is that a 
virialized halo has a memory of its formation history in the sense 
that at each formation step the total energy of infalling matter is conserved.
To determine the virialized structure, two additional simplifying assumptions 
have to be made. We suppose that a) the dynamical time-scale of 
the pre-existing structure is short compared to the time-scale of 
the infall (i.e. the pre-existing structure is virialized) 
and b) that the structure already present reacts adiabatically 
to the infall. In detail, the three assumptions about infall 
of matter at formation step $k$ are:
\begin{itemize}
\item[i)] {\it Total Energy}: The total energy of the infalling mass 
$m_k$ is conserved. There is no energy dissipation and 
\begin{equation}
E_k = const.
\end{equation}
\item[ii)] {\it Virialization}: The dynamical time-scale of the pre-existing
structure is short compared to the time-scale of the infall. The infalling mass
$m_k$ virializes in the gravitational potential originating from the pre-existing, 
virialized structure. The corresponding virial theorem is 
(e.g. Binney \&Tremaine 1987)
\begin{equation}
2 K_k + W_k + V_k =0 
\label{virial}
\end{equation}
where $K_k$ is the kinetic, $W_k$ the potential energy and $V_k$ is related to
the external gravitational potential.
\item[iii)] {\it Adiabatic Contraction}: The pre-existing structure reacts
adiabatically to the infalling matter deposited within its extension. 
This assumption is based on the fact that for periodic orbits $\oint pdq$ 
(where $p$ and $q$ are then conjugate momentum and coordinate, respectively)
is an adiabatic invariant (Blumenthal et al. 1986, Flores et al. 1993).
\end{itemize} 
To simplify calculations, the density distribution resulting from
virialization of the mass $m_k$ is approximated by  
\begin{equation}
\bar{\rho}_k(r)=
\left\{ \begin{array}{cc} 
 {3 {m_k /{4 \pi R_k^3}}}& r\le R_k  \\
   0			  & r> R_k 
\end{array}  \right.
\label{densdist}
\end{equation}
where $R_k$ is the characteristic (virial) radius determined by (\ref{virial}).

Before discussing assumption i), i) and iii) in more detail, consider 
the sequence of formation steps. The formation of the virialized halo 
starts with the collapse of the mass $m_K$ at redshift $z_K$. 
There is no pre-existing structure and the collapse according to the three
assumptions i), ii) and iii) is equivalent to the familiar top-hat collapse.
The resulting object (structure $K$) is characterized by the density
distribution $\bar{\rho}_K$ (\ref{densdist}) where $R_k$ is determined from 
(\ref{virial}) with $V_K=0$. 

The mass $m_{K-1}$ collapses onto structure $K$ at redshift $z_{K-1}$. 
Its virialized density distribution $\bar{\rho}_{K-1}$ is determined by 
eqs. (\ref{virial},\ref{densdist}) depending on $\bar{\rho}_K$. The mass 
enhancement within $R_{K}$ due to the infall of $m_{K-1}$ leads to an 
adiabatic contraction of the innermost structure $K$ (assumption iii). 
It is estimated here by considering the quantity $R_K M_K$ as adiabatic 
invariant. In principle, this implies a self-similar mass distribution. 
Since we do not attempt to spatially resolve $\bar{\rho}_K$ within $R_K$, 
this is a reasonable approximation here. The contraction of structure $K$ 
due to the infall of $m_{K-1}$ then is described by
\begin{equation}
[m_{K-1}(R'_{K})+m_K]R'_{K}=m_K R_K
\label{contraction}
\end{equation} 
where $m_{K-1}(R'_{K})$ is the fraction of the mass $m_{K-1}$ which is inside 
the 'contracted' radius $R'_{K}$.

In formation step $K-2$, the density distribution  
$\bar{\rho}_{K-2}$ is determined by eqs. (\ref{virial},\ref{densdist}) depending 
now on $\bar{\rho}_K$ and  $\bar{\rho}_{K-1}$. The infall of mass $m_{K-2}$ 
leads to a contraction of structure $K-1$ and the structure $K$ contracts because
of $m_{K-2}$ {\it and} the contraction of structure
$K-1$ (which contributes to a mass enhancement within $R_K$ as well). 

The formation of the collapsed object is completed at $z=0$ after the
formation steps $K-3,K-4,\ldots,0$. The structure of the 
virialized halo is described as a superposition of concentric spheres 
of different but uniform density and defined by
the set $\{(m_k,R_k), k=0,1,2,\ldots,K\}$. The total density profile is 
\begin{equation}
\rho(r)=\sum_{k=0}^K \bar{\rho}_{k}(r).
\label{rhotot}
\end{equation}

Let us stress similarities and differences between this model and earlier
work. With respect to the assumption i) and  iii) (no energy dissipation, 
adiabatic response), the model is equivalent to the collapse model of 
Gunn (1977) and models (e.g. Hoffman \& Shaham 1985, Hoffman 1988) 
based on the stable clustering assumption (Filmore \& Goldreich 1984, 
Bertschinger 1985). The similarity partially extends to the process 
of virialization in the sense that a stepwise, separated virialization 
of infalling matter is assumed. However there are two main differences: 
The model makes no assumptions about self-similarity and it includes 
the formation history. With respect to these differences, the presented 
model clearly adopts a more realistic view of hierarchical structure formation. 

But how realistic are assumptions i)-iii) in the context of hierarchical 
structure formation? A complete answer to this question lies beyond the 
scope of this letter. The key issue probably is assumption i).
%, i.e. conservation of the total energy of infalling matter. An 
attempt to semi-analytically model heating of the pre-existing structure
by infalling matter has recently been made by Nusser \& Sheth (1998). 
Their study seems to indicate that including energy dissipation (e.g.
by drag forces, Ostriker \& Turner 1979) does not drastically alter density 
profiles but leads to a slightly steeper slope in the innermost region of 
halos. Regarding assumption ii) the time-scale for virialization which is 
comparable to the dynamical time-scale 
$\tau_{vir} \sim (4 \pi G \rho_{max})^{-1/2}$ is of importance. As the 
virialized structure forms and the cuspy inner region emerges, $\tau_{vir}$ 
rapidly gets smaller. From the outcome of the model calculation (c.f. 
Section \ref{results}) one finds that for timestep 
$\tau_{in} \simeq \Delta z \sim 0.05$ and after
the first few formation steps $\tau_{vir}\lesssim\tau_{in}$. 
This indicates that it is reasonable to assume infalling matter to collapse 
onto an at least pre-virialized structure. Finally, it should be stressed
that the model uses energy considerations to relate mass histories and
virialized density distributions. 
These considerations depend on the gravitational potential which itself 
is much smoother and more symmetric than the underlying density distribution 
(see also Hoffman 1987).

\section{Results}\label{results}
The results presented here are based on $50$ Monte Carlo 
realizations of the formation history of halos of mass 
$M_0=10^{11} M_\odot$, $10^{13} M_\odot$ and $10^{15} M_\odot$ 
in the context of the SCDM cosmology ($\Omega=1,h=0.5,\sigma_8=0.67$). 
A timestep $\Delta z=0.05$ has been chosen and the mass limit below which 
the object is considered to be a seeding object is defined 
as $M_f\le0.05 M_0$. While the time step $\Delta z$ affects the 
mean mass falling in at each formation step, the resulting virialized
density distributions do not change significantly in the tested
range $\Delta z=0.01 \ldots 0.2$.

\subsection{Virialized Density Distribution}
\begin{figure*}[h]
\epsscale{1.0}
\plotone{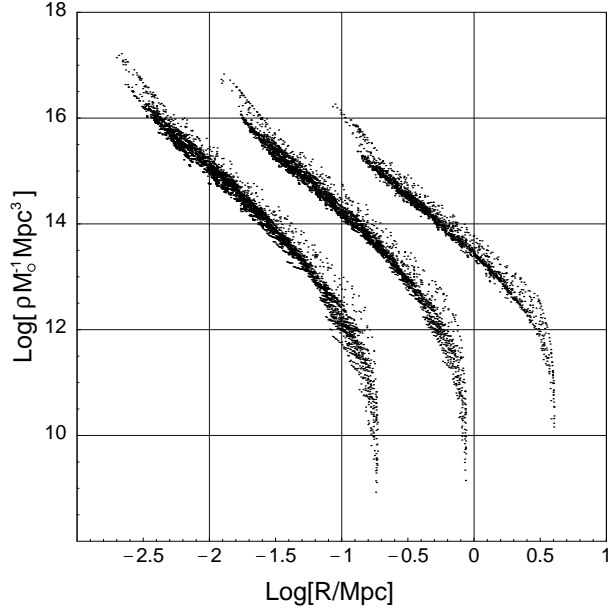}
\caption{\label{fig1} Superposition of the virialized density distributions resulting 
from the $50$ mass histories and the collapse model discussed in Section 
\ref{model}. The halo masses are $10^{11} M_\odot$, $10^{13} M_\odot$ and 
$10^{15} M_\odot$ (from left to right).}
\end{figure*}

Figure \ref{fig1} shows a superposition of the virialized density 
distributions. Data points mark the total density at the 
various radii $R_k$. The sharply dropping density at the outer
edge of the distributions originates from the missing infall of
mass in excess of the defined halo mass.
The radius at which this drop occurs roughly corresponds to 
the radius $r_{200}$ at which the mean overdensity drops 
below $\delta=200$. The radius $r_{200}$ is known 
as the characteristic size of a virialized object. This implies 
that for $r>r_{200}$ the assumptions of the model break down. As a 
consequence, the density distribution at $r>r_{200}$ will not be 
considered hereafter. Table \ref{tab2} presents the mean of $r_{200}$
and the corresponding mass $M_{200}$. The scatter of the innermost 
data points is due to the different formation masses $M_f$ and 
redshifts $z_f$ leading to different initial values of $R_1$ 
and $\bar{\rho}_1$, respectively. For radii $r<R_1$, the 
structure of the density distributions are not resolved. Thus the 
meaningful innermost radius for an individual density distribution 
is $r_{\mbox{\tiny res}}=R_1$. The radial range of the model thus 
is $[r_{\mbox{\tiny res}},r_{200}]$.  

\begin{figure*}[h]
\epsscale{1.0}
\plotone{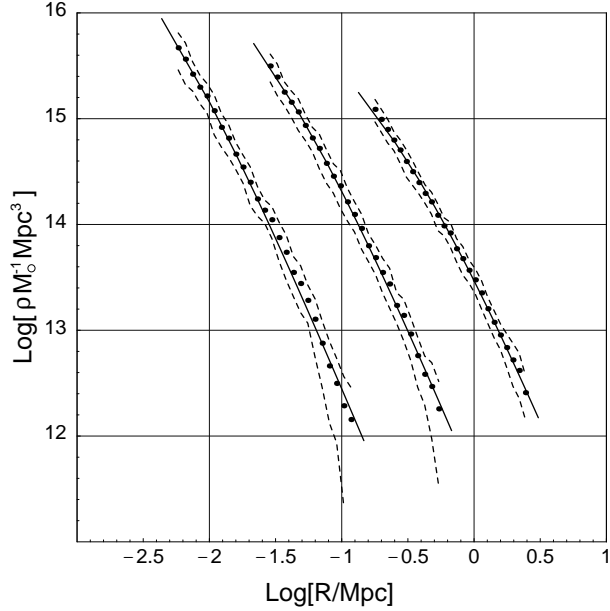}
\caption{\label{fig2} Mean virialized density distribution (points) and rms scatter 
(dashed lines). The solid line shows the best fitting NFW profile. The outermost 
data points mark $r=r_{200}$. For a discussion of the inner limit c.f. text.}
\end{figure*}

\begin{figure*}[h]
\epsscale{1.0}
\plotone{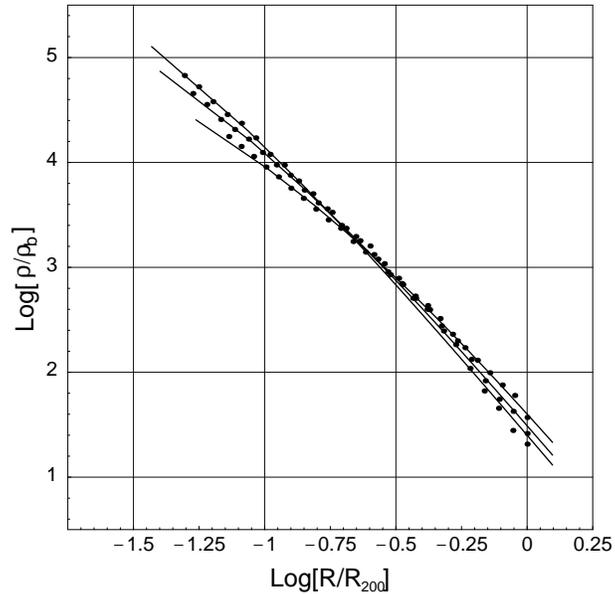}
\caption{\label{fig3} Scaled mean density profiles (points) and the best fitting NFW 
profiles of the $10^{11} M_\odot$, $10^{13} M_\odot$ and $10^{15} M_\odot$ mass halos
(from top to bottom at $r/r_{200}<0.1$). For better readability, plotting of the 
rms scatter has been omitted. The scaled profiles agree within one $\sigma$ 
for $r/r_{200}>0.1.$ Note the different behavior of the density distributions 
for $r/r_{200}<0.1$. Higher halo masses imply less centrally concentrated halos.}
\end{figure*}

Figure \ref{fig2} shows the mean virialized density profile. Data 
points mark the mean of the $50$ individual density profiles for the 
$10^{11} M_\odot$, $10^{13} M_\odot$, and $10^{15} M_\odot$ halos 
(from left to right). Dashed lines indicate the rms scatter. The data 
points mark the radial range $[\bar{r}_{\mbox{\tiny res}},\bar{r}_{200}]$. The 
best fit NFW profile (\ref{nfwprof}) is plotted as solid line. Table \ref{tab2} 
lists the concentration parameter $r_0/r_{200}$.  Higher halo masses lead 
to less centrally concentrated halos. The comparison of Figures \ref{fig2} 
and \ref{fig3} with Figures 3 and 4 of NFW96 and corresponding 
concentration parameters illustrates the agreement with N-body simulations.

\subsection{Circular Velocities}
Mean circular velocities and the corresponding scatter are calculated 
from $v_c(r)=\sqrt{G M(r)/{r}}$. The maximum values for the scatter are 
found roughly at $r=r_{\mbox{\tiny res}}$.
It is $\pm13\%$, $\pm10\%$ and $\pm6\%$ for the $10^{11}M_\odot$ $10^{13}M_\odot$ 
and $10^{15}M_\odot$ halos, respectively. These values are in good agreement with 
Eisenstein \& Loeb (1995) (see Figure 2 in their paper). For the outer regions 
of the halos, the model predicts a significantly decreasing scatter which 
reaches values well below $10\%$ at $r=r_{200}$ (c.f. Table \ref{tab2}). 
If confirmed by N-body simulations, this could be of 
importance for the interpretation of the Tully-Fisher, e.g. in terms of
its relation to cosmological initial conditions (see Eisenstein \& Loeb 1995).
As a word of caution it should be noted that the poor radial density resolution 
at $r \lesssim r_{\mbox{\tiny res}}$ affects the circular rotation profiles. 
Compared to the NFW profile, the present model suggests a higher central 
density leading to flatter circular velocity profiles. 

\section{Summary and Conclusions\label{discussion}}
A simple analytic collapse model has been proposed for the hierarchical 
formation scenario of virialized cold dark matter halos. As a working 
hypothesis it has been assumed that dark matter halos posses a memory 
of their formation history by conserving the total energy of infalling 
matter in each formation step. For the tested SCDM cosmology, the virialized 
mean density profiles are in good agreement with numerical results. 
This suggests that energy dissipation does not play a major role in the 
determination of the mean shape of virialized density profiles. As a 
consequence, the (nonlinear) virialized density distribution is tightly 
related to the (linear) density field of the infall region by the total 
energy of infalling matter.  

An improved model could include more realistic density distributions $\bar{\rho}_k$
and probably should differentiate between minor and major mergers 
(e.g. Salvador-Sol\'e et al. 1998). While these modifications could allow for 
resolving the radial range $r<0.05 r_{200}$, it is believed that they will not 
drastically alter the results for $r>0.05 r_{200}$.  

\acknowledgments
Stimulating discussions with L. Kofman and J. R. Bond
are gratefully acknowledged. This work was supported by the Swiss 
National Foundation, grant 81AN-052101.

\begin{deluxetable}{ccccccc}
\label{tab:mass}
\tablewidth{35pc}
\tablecaption{Parameters of the Mean Virialized Density Profiles}
\tablehead{
\colhead{$M_0$} &\colhead{$M_{200}$} & \colhead{$r_{200}$} & \colhead{$V_{200}$} & 
\colhead{$\rho_0$} & \colhead{$r_0$} & \colhead{$r_0/r_{200}$} \nl
[$M_\odot$]  &[$M_0$] & [Mpc] & [km s$^{-1}$] & [M$_\odot$ Mpc$^{-3}$] & [Mpc] & }
\startdata
$10^{11}$ & $0.94\pm0.05$ & $0.117\pm 0.002$  &  $58.7 \pm 0.97$ & $3.25 \times 10^{16}$ & 0.0044 & $0.038$\nl
$10^{13}$ & $0.92\pm0.05$ & $0.541\pm 0.009$  &  $270\pm 4.6$ & $7.33 \times 10^{15}$ & 0.0373 & $0.069$\nl
$10^{15}$ & $0.85\pm0.05$ & $2.44\pm 0.05$    &  $1222\pm 25$ &$1.37 \times10^{15}$ & 0.3367 & $0.138$\nl
\enddata
%\tablecomments{\label{tab2}}
\label{tab2}
\end{deluxetable}

\end{document}